# Measuring scientific output of researchers by *t*-index and Data Envelopment Analysis

Dušan Teodorović[1,2,*]    Miloš Nikolić[2]

[1]Serbian Academy of Sciences and Arts, Knez Mihailova 35, 11000 Belgrade, Serbia
[2]University of Belgrade, Faculty of Transport and Traffic Engineering, Vojvode Stepe 305, 11000 Belgrade, Serbia

**Abstract**

There is a growing need for ranking universities, departments, research groups, and individual scholars. Usually, the scientific community measures the scientific merits of the researchers by using a variety of indicators that take into account both the productivity of scholars and the impact of their publications. We propose the t-index, the new indicator to measure the scientific merits of the individual researchers. The proposed t-index takes into account the number of citations, number of coauthors on every published paper, and career duration. The t-index makes the possible comparison of researchers at various stages of their careers. We also use in this paper the Data Envelopment Analysis (DEA) to measure the scientific merits of the individual researchers within the observed group of researchers. We chose 15 scholars in the scientific area of transportation engineering and measured their t-index values, as well as DEA scores.

1. Introduction

Education and research form the world of the future. It is well-known and widely accepted, that education and research are the key issues for development, economic growth, new knowledge, new technologies, and social and cultural transformation. According to the "Universities Worldwide" (https://univ.cc/states.php) there are 9,548 universities in 207 countries. As stated by World Economic Forum, 67,449 students graduated with a Ph.D. in the US in 2014. In the same year, Germany had 28,147 Ph.D. graduates, the United Kingdom had 25,020, Japan had 16,038, etc. Thousands of Ph.D. students, researchers, and professors read, study, think, write, develop mathematical models, perform biological and chemical experiments, test drugs, create patents, write computer codes, perform clinical trials, achieve more or less significant scientific results and publish scientific papers. There are numerous decisions at many world universities, related to candidate ranking, candidate promotion, tenure position, research proposal acceptance that highly depend on the researcher's scientific merits. The research community is obliged, on a daily basis, to measure the scientific merits of the researchers.

---

[*] Corresponding author
e-mail addresses: duteodor@gmail.com (Dušan Teodorović), m.nikolic@sf.bg.ac.rs (Miloš Nikolić)

At many world universities, young researchers are subjected to great pressure to publish papers. Such practice often leads to a decline in research quality. As mentioned by Garousi and Fernandes (2017), "more than 70,000 papers have been published in the area of Software Engineering (SE) since the late 1960s. The 43% of those papers have received no citations at all." Recognizing the fact that extensive publishing is often at the expense of top research, many state agencies and universities are increasingly promoting the policy of excellence. Research Councils UK defines academic impact as "the demonstrable contribution that research makes to academic advances, across and within disciplines, including significant advances in understanding, methods, theory and application." There is an increasing need for ranking universities, departments, research groups, and individual scholars. Traditionally, the scientific community measures the scientific merits of the researchers by using various indicators that take into account both the productivity of researchers and the impact of their publications. The widely accepted indicator is the *h*-index, proposed by Hirsch (2005). Several other indicators have been also suggested in the literature (*g*-index (Egghe 2006), Hirsch core (Rousseau 2006), *A*-index (Jin 2006), *R*-index (Jin 2007), h(2)-index (Kosmulski 2006), *AR*-index (Jin 2007).

In this paper, we propose a new indicator to measure the scientific merits of individual researchers. We call the proposed indicator the *t*-index. The proposed *t*-index takes into account the number of citations, number of authors on every published paper, and career duration. The proposed *t*-index makes the possible comparison of researchers at various stages of their careers.

We also use in the paper the Data Envelopment Analysis (DEA) to measure the scientific merits of the individual researchers *within* the observed group of researchers. Each individual researcher denotes the entity that changes inputs into outputs. The DEA defines the efficiency for every researcher as a weighted sum of outputs divided by a weighted sum of inputs. The efficiency of any researcher is within the range [0, 1], or [0%, 100%].

## 2. The *t*-index

Several indices try to measure the scientific merits of individual scholars. Practically, all of them are based on researcher citations that "are widely recognized as a standard measure of academic influence" (Van Noorden 2017). Within citation analysis, we count how many times scholars' published papers are cited. The basic premise of the citation analysis is that significant scientific contributions are cited more than less significant.

The generally accepted indicator is the *h*-index. This index was proposed by Hirsch (2005) who defined it in the following way: "A scientist has index $h$ if $h$ of his or her $N_p$ papers have at least $h$ citations each and the other $N_p - h$ papers have $\leq h$ citations each." The *h*- index takes into account the number of scholar's publications and their impact on the scientific community. With the intention of decreasing the effects of co-authorship, Batista et al. (2006) introduced Individual *h*-index by dividing the *h*-index by the average number of authors of the papers that contribute to the *h*-index value. Before the appearance of the *h*-index, the indicators based on a single criteria (total number of published papers, total number of citations) were primarily used in the process of individual scholar evaluation. All main scientific databases (*Web of Science*, *Scopus*, *Google Scholar* and *Publish or Perish*) calculate and show the *h*-index for every scholar. These databases have different coverage of the scientific literature and consequently produce different results that describe the impact of individual scholar's work and scientific results.

Egghe (2006) offered to the scientific community the *g*-index. This index represents a modification of the *h*-index and is defined in the following way: "The highest number *g* of papers that together received $g^2$ or more citations." Regardless of the "publish or perish" atmosphere, in which the world academic community lives, there are a significant number of scholars who publish papers only in the case of very significant scientific results. As a consequence, these scholars have a small number of published papers that are highly cited, as well as low values of *h*, and *g* indices. On the other hand, the good thing about the *h*-index is the fact that it cannot be improved by producing a big number of lowly cited papers.

Rousseau (2006) and Jin et al. (2007) proposed the *A*-index and defined it as follows:

$$A = \frac{1}{h}\sum_{j=1}^{h} c_j \qquad (1)$$

where:

*h* - the *h*-index
$c_j$ - the number of citations of the *j*-th most cited paper

Jin et al. (2007) proposed the *R*-index which is defined in the following way:

$$R = \sqrt{\sum_{j=1}^{h} c_j} \qquad (2)$$

Practically, both *A* and *R* indices try to measure the citation intensity of the *h* most cited scholar's papers.

Kreiman and Maunsell (2011) proposed "criteria that a metric should have to be considered a good index of scientific output". They stated that the indicator should be quantitative, and based on data that are rapidly updated and retrospective. Kreiman and Maunsell (2011) elaborated that the number of contributors should be normalized, as well as that normalization should be performed by scientific area, and for career stage.

We agree with recommendations made by Kreiman and Maunsell (2011). By proposing the *t*-index we try to propose a very simple metric capable to take into account the number of co-authors, and career stage. Let us assume that we have to evaluate and rank researchers from the same research area. We introduce the following notation:

$R$ - total number of researchers to be ranked

$i = 1,2,…, R$  - researcher index

$P_i$ - total number of papers of the *i*-th researcher

$y_i$ - number of years since the first paper of the *i*-th researcher appeared

$c_{ij}$ - number of citations of the *j*-th paper of the *i*-th researcher

$C_i = \sum_{j=1}^{P_i} c_{ij}$ - total number of citations of papers on which the *i*-th researcher appears as one of the authors

$n_{ij}$ – number of authors of the *j*-th paper of the *i*-th researcher

The total number of published papers $P_i$ and the total number of scholar's citations $C_i$ represent two basic indicators of the scholar's productivity and the impact of his/her research. We do believe that number of citations of every published paper $c_{ij}$ should be divided by the total number of authors $n_{ij}$ of the paper. In this way, we prevent multiple counting of citations, and try to prevent false authorship. In other words, we use the following sum to represent the scientific impact $SI_i$ of the scholar:

$$SI_i = \sum_{j=1}^{P_i} \frac{c_{ij}}{n_{ij}} \quad i = 1,2,…,R \qquad (3)$$

Certainly, there are differences in terms of the usual number of co-authors in natural sciences, mathematics, medicine, engineering, or biotechnical sciences. Depending on the theoretical or experimental research and/or scientific area, the scientific impact of the scholar $SI_i$ could be also described in the following way:

$$SI_i = \sum_{j=1}^{P_i} \frac{c_{ij}}{1+a \cdot (n_{ij}-b)} \qquad for\ n_{ij} > b \qquad (4)$$

where:

$a$ – penalty that prevents big number of co-authors

$b$ - usual number of co-authors in the considered scientific discipline

In order to be able to make comparison of scholars at various stages of their career, we introduce into the analysis the number of years $y_i$, since the first paper of the *i*-th researcher appeared.

We measure the scientific merits of the individual scholar by the *t*-index that we define in the following way:

$$t_i = \frac{SI_i}{y_i} \qquad i = 1,2,\ldots,R \tag{5}$$

i.e.:

$$t_i = \frac{\sum_{j=1}^{P_i} \frac{c_{ij}}{n_{ij}}}{y_i} \qquad i = 1,2,\ldots,R \tag{6}$$

To have high regard for scientists that try to concentrate on a more significant, more serious, and long-lasting research, instead of publishing papers that attract little attention of the academic community, it is possible to take into account only researcher's papers that have a significant citation. In other words, we can introduce the threshold $c^*$ that represents minimal number of citations that qualifies paper to be included in the citation analysis. Any paper, that has number of citations greater than or equal to $c^*$, should be included in the analysis. The *t*-index defined in this way, which focuses on research quality over quantity, reads:

$$t_i = \frac{\sum_{j=1}^{P_i} \frac{\delta_{ij} \cdot c_{ij}}{n_{ij}}}{y_i} \qquad i = 1,2,\ldots,R \tag{7}$$

where:

$$\delta_{ij} = \begin{cases} 1, & if\ c_{ij} \geq c^* \\ 0, & otherwise \end{cases} \tag{8}$$

Obviously, the threshold $c^*$ must be different for various scientific disciplines. There are significant inter-field differences in the usual number of published papers, as well as in the total number of citations of individual researchers. Consequently, the *t*-index cannot be applied, without adequate normalization, to compare scientists from different scientific areas.

### 3. Measuring scientific output of individual researchers within the group by the Data Envelopment Analysis

Frequently there is a need to measure the scientific achievements of the individual researchers *within* the group of researchers. The *Data Envelopment Analysis* (DEA) is a measurement technique that is applied for evaluating the relative efficiency of decision-making units (DMU's). The DMU could be countries, regions, cities, airports, hospitals, universities, firms, etc. The DEA is also called *frontier analysis*. By using this technique, the analyst could estimate the efficiency of any number of DMU's. The studied DMU's could have any number of inputs and outputs. Various DEA models and applications have been proposed during the last four decades. In this paper, we treat individual scholars as decision-making units.

The DEA was originally proposed by Charnes et al. (1978). Each DMU indicates the entity that transforms inputs into outputs. The DEA defines the relative efficiency in the following way:

$$Efficiency = \frac{Weighted\ sum\ of\ outputs}{Weighted\ sum\ of\ inputs} \tag{9}$$

$$Efficiency = \frac{u_1 \cdot y_{1j} + u_2 \cdot y_{2j} + \cdots + u_n \cdot y_{nj}}{v_1 \cdot x_{1j} + v_2 \cdot x_{2j} + \cdots + v_m \cdot x_{mj}} \tag{10}$$

where:

$u_i$ - weight of the output $i$
$y_{ij}$ - amount of output $i$ from the unit $j$
$x_{ij}$ – amount of input $i$ to the unit $j$
$v_i$ - weight of the input $i$

The DEA calculates the efficiency for every DMU as a weighted sum of outputs divided by a weighted sum of inputs. The efficiency of any DMU is inside the range [0, 1], or [0%, 100%]. When calculating the efficiency of a specific DMU by the DEA technique, the weights of a DMU are chosen to present the considered DMU in the best possible light.

The efficiency $h_0$ of the DMU $j_0$ is calculated by solving the following fractional programming problem:

Maximize

$$h_0 = \frac{\sum_r u_r \cdot y_{rj_0}}{\sum_i v_i \cdot x_{ij_0}} \quad \forall j \tag{11}$$

subject to:

$$\frac{\sum_r u_r \cdot y_{rj_0}}{\sum_i v_i \cdot x_{ij_0}} \leq 1 \tag{12}$$

$$u_r, v_i \geq \varepsilon \tag{13}$$

By solving the fractional programming problem, the analyst obtains the input weights $v_i$, as well as the output weights $u_r$. In the case of $n$ DMU to be evaluated, the analyst has to perform $n$ optimizations (one for every DMU to be evaluated). The fractional program could be replaced by the following equivalent linear program:

Maximize

$$h_o = \sum_r u_r \cdot y_{rj_o} \quad (14)$$

subject to:

$$\sum_i v_i \cdot x_{ij_o} = 1 \quad (15)$$

$$\sum_r u_r \cdot y_{rj} \leq \sum_i v_i \cdot x_{ij} \quad \forall j \quad (16)$$

$$u_r, v_i \geq \varepsilon \quad (17)$$

The inputs and outputs in the DEA have various units. The DEA enables the analyst to directly compare considered DMU with their peers.

DEA has been extensively used to measure the efficiency of universities. In their pioneering work, Tomkins and Green (1988) used the DEA to evaluate the efficiency of UK University Departments of Accounting. Johnes and Johns (1993), Stevens (2005), and Thanassoulis et al. (2011) measured the research performance of UK economics departments. Beasley (1995) developed a DEA-based model for jointly determining teaching and research efficiencies for university departments. Athanassopoulos and Shale (1997) and Flegg et al. (2004) measured the efficiency of the UK, while Madden et al. (1997), Avkiran (2001), and Abbott and Doucouliagos (2003) measured the efficiency of Australian universities by the DEA. Rousseau, S. and Rousseau, R. (1997) measured the research efficiency of the developed countries. In their DEA model, they used GDP, active population, and R&D expenditure as inputs and the number of publications in the SCI journals, and the number of patents granted by the European Patent Office as outputs. McMillan and Datta (1998) measured the relative efficiencies of Canadian universities by the DEA. Hanke and Leopoldseder (1998) used DEA to compare the efficiency of Austrian universities. Abramo and D'Angelo (2009) and Abramo et al. (2011) used the Data Envelopment Analysis (DEA) to evaluate the efficiency of research activities of the Italian university system, while Taylor and Harris (2004) compared South African universities by the DEA.

Petridis et al. (2013) evaluated forestry scientific journals by using Data Envelopment Analysis (DEA). The authors used the frequency of publication of a journal within a year, and the number of articles published per year as inputs. The outputs were the Eigenfactor score, $h$-index, and 5-year impact factor.

Agrell and Steuer (2000) developed the DEA-based model called ACADEA to evaluate individual performances of the faculty. They used research output, teaching output, external service, internal service, and cost to evaluate individual faculty. The authors applied the developed system to a university department with 30 faculty members that were evaluated over a 3-year period. DEA is not frequently used to evaluate the performances of individuals in the academic community. The paper of Agrell and Steuer (2000) is definitely one of the pioneering papers in the teaching and research evaluation of individual faculties.

The individual scientific achievements are the result of knowledge, talent, dedication, enthusiasm, number of co-authors, equipment available to researchers, etc. Obviously, it is not possible to include all these factors in the process of measuring the scientific output of individual scholars. We decide to measure the scientific output of individual researchers by using the following inputs and outputs (Figure 1):

Input 1: $y_i$ - number of years since the first paper of the $i$-th researcher appeared

Input 2: $N_i = \sum_{j=1}^{P_i} n_{ij}$ - total number of coauthors of the $i$-th researcher

Output: $C_i = \sum_{j=1}^{P_i} c_{ij}$ - total number of citations of the $i$-th researcher papers

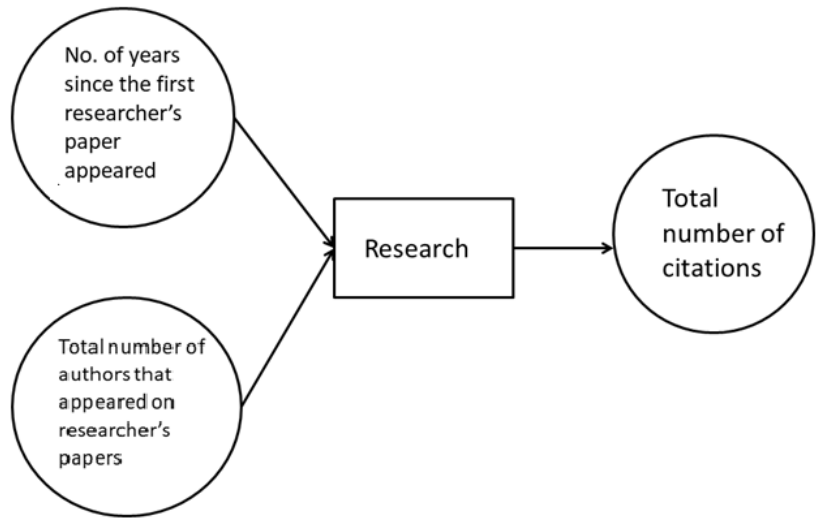

Figure 1. Measuring the scientific output of individual researchers

The chosen inputs - number of years since the first paper of the researcher appeared, and the total number of researcher's coauthors, and the output - total number of researcher's citations could be obtained from the existing basis - *Google Scholar*, *Web of Science* and *Scopus*. These inputs are also highly connected with the scientific impact of the individual researcher.

## 4. Example

We approached the base Google Scholar in June 2018. We chose 15 scholars in the scientific area of transportation engineering. The number of years since the first paper of the researcher appeared, the total number of authors that appeared on the researcher's papers, the total number of researcher's citations, and the researcher's $h$-index is given in Table 1.

Table 1 - The number of years since the first paper of the researcher appeared, the total number of researcher's coauthors, total number of researcher's citations, and researcher's $h$-index.

| Researcher | Number of years since the first researcher's paper appeared | Total number of authors that appeared on researcher's papers | Total number of citations | $h$-index |
|---|---|---|---|---|
| $R_1$ | 40 | 350 | 5977 | 39 |
| $R_2$ | 7 | 32 | 193 | 6 |
| $R_3$ | 45 | 475 | 3359 | 31 |
| $R_4$ | 31 | 300 | 1343 | 18 |
| $R_5$ | 21 | 121 | 964 | 17 |
| $R_6$ | 52 | 398 | 8480 | 43 |
| $R_7$ | 39 | 1127 | 16276 | 62 |
| $R_8$ | 35 | 208 | 1642 | 24 |
| $R_9$ | 27 | 149 | 1812 | 17 |
| $R_{10}$ | 13 | 129 | 1798 | 18 |
| $R_{11}$ | 15 | 123 | 837 | 12 |
| $R_{12}$ | 9 | 141 | 319 | 8 |
| $R_{13}$ | 35 | 400 | 1184 | 16 |
| $R_{14}$ | 9 | 108 | 153 | 8 |
| $R_{15}$ | 22 | 179 | 467 | 13 |

The calculated values of the $t$-index and efficiency scores obtained by the DEA analysis are given in Table 2.

Input resources (number of authors and time spent in research) utilized by observed scholars to produce one citation are graphically shown in Figure 2. We determine the efficient frontier by selecting the scholars for which one input/output cannot be improved without worsening the other input/output (Figure 2).

Table 2. The calculated values of the *t*-index and efficiency scores obtained by the DEA analysis

| Researcher | $t$ - index | Efficiency (DEA CCR) |
|---|---|---|
| $R_1$ | 79.826 | 0.848 |
| $R_2$ | 12.702 | 0.283 |
| $R_3$ | 33.077 | 0.377 |
| $R_4$ | 13.874 | 0.231 |
| $R_5$ | 16.893 | 0.374 |
| $R_6$ | 66.345 | 1.000 |
| $R_7$ | 180.028 | 1.000 |
| $R_8$ | 19.630 | 0.370 |
| $R_9$ | 30.825 | 0.570 |
| $R_{10}$ | 61.618 | 0.727 |
| $R_{11}$ | 16.766 | 0.329 |
| $R_{12}$ | 17.486 | 0.137 |
| $R_{13}$ | 12.259 | 0.162 |
| $R_{14}$ | 4.738 | 0.079 |
| $R_{15}$ | 6.886 | 0.126 |

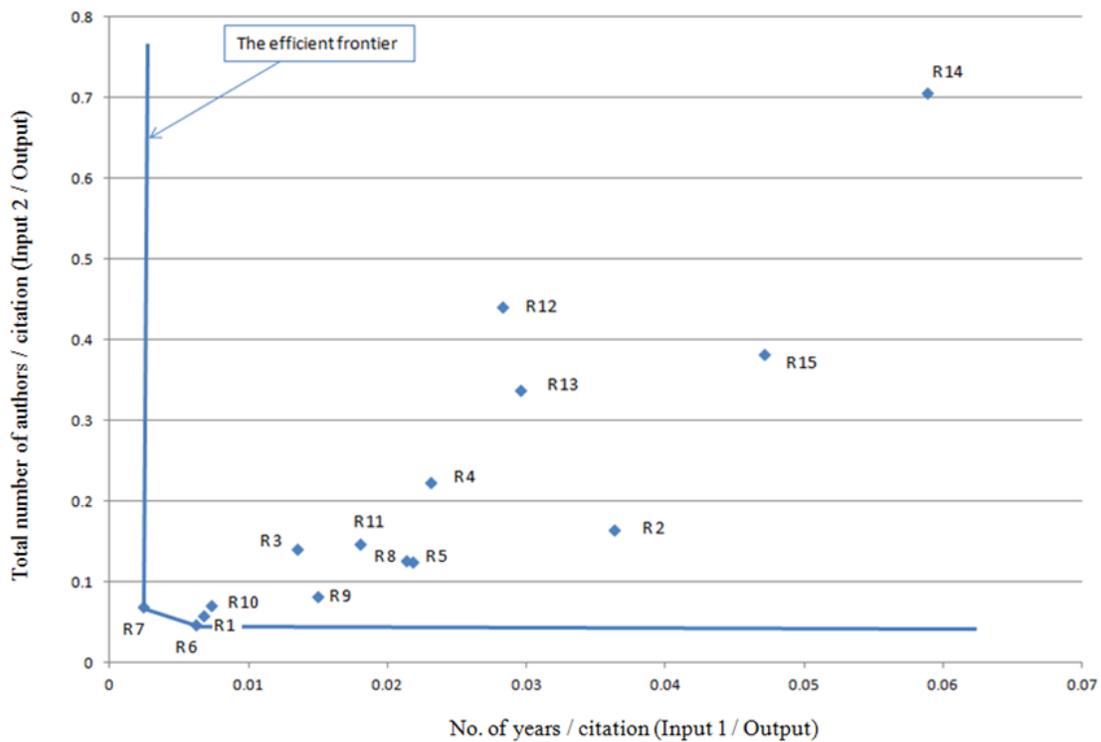

Figure 2. Input resources utilized by observed scholars to produce one citation

Table 3 shows the scholar ranks according to *t*-index, DEA, and *h*-index.

Table 3. The scholar ranks according to *t*-index, DEA, and *h*-index

| Researcher | Rank according to *t*-index | Rank according to DEA | Rank according to *h*-index |
|---|---|---|---|
| $R_1$ | 2 | 3 | 3 |
| $R_2$ | 12 | 10 | 15 |
| $R_3$ | 5 | 6 | 4 |
| $R_4$ | 11 | 11 | 6 |
| $R_5$ | 9 | 7 | 8 |
| $R_6$ | 3 | 1 | 2 |
| $R_7$ | 1 | 1 | 1 |
| $R_8$ | 7 | 8 | 5 |
| $R_9$ | 6 | 5 | 8 |
| $R_{10}$ | 4 | 4 | 6 |
| $R_{11}$ | 10 | 9 | 12 |
| $R_{12}$ | 8 | 13 | 13 |
| $R_{13}$ | 13 | 12 | 10 |
| $R_{14}$ | 15 | 15 | 13 |
| $R_{15}$ | 14 | 14 | 11 |

We calculated correlation coefficients to see how close are the scholar ranks according to *t*-index, DEA, and *t*-index. We obtained the following correlation coefficient values:

$R_{t,DEA} = 0.93$
$R_{t,h} = 0.82$
$R_{t,A} = 0.86$
$R_{DEA,h} = 0.82$
$R_{DEA,A} = 0.87$

Both, the *t*-index and the DEA take into account the number of citations, the number of coauthors on every published paper, and career duration. On the other hand, *the h*-index takes care exclusively of individual scholar's achievements, ignoring the number of coauthors and career duration. Consequently, the correlation coefficient is the highest between the *t*-index values and DEA efficiency scores. It is important to note that the values of the correlation coefficient are also high between the *t* and the *h* -index, as well as between the DEA scores and the *h*-index. It would be important to determine the values of the correlation coefficient for a significantly larger sample of researchers.

We also measured observed researchers based on researcher's papers that have significant citation. We introduced the threshold $c^*=50$. All papers that had number of citations greater than or equal to 50 were included in the analysis. The obtained $t_{50}$ - index values are given in Table 4.

Table 4. The $t_{50}$ - index values

| Researcher | $t$ - index | $t_{50}$ - index |
|---|---|---|
| $R_1$ | 79.8255 | 64.229 |
| $R_2$ | 14.8194 | 10.333 |
| $R_3$ | 33.8284 | 17.061 |
| $R_4$ | 14.3367 | 6.013 |
| $R_5$ | 16.8929 | 7.349 |
| $R_6$ | 67.6462 | 56.653 |
| $R_7$ | 180.028 | 149.307 |
| $R_8$ | 20.2074 | 7.02 |
| $R_9$ | 29.7244 | 21.202 |
| $R_{10}$ | 72.8212 | 60.674 |
| $R_{11}$ | 9.3142 | 5.178 |
| $R_{12}$ | 19.6714 | 14 |
| $R_{13}$ | 11.9185 | 1.289 |
| $R_{14}$ | 5.33006 | 0 |
| $R_{15}$ | 9.46801 | 1.547 |

The correlation coefficient between $t$ - index and $t_{50}$ - index equals $R_{t,t50} = 0.957$. In other words, this preliminary analysis shows that the $t$-index highly depends on the most cited papers of the researcher.

## 5. Conclusion

The evaluation of the scientific results of the researchers is a complex and responsible task. Certainly, it is not possible to describe, with a single numerical indicator, a scholar's productivity, and scientific impact. On the other hand, numerical indicators are of great help in the process of individual scientist evaluation.

The number of citations is currently the basic input to measure the individual scientist's success. Practically, all the proposed indexes that try to measure an individual researcher's scientific output have been based on the number of citations.

The new indicator that we propose, to measure the scientific achievements of the individual scholars, is no exception in this sense. We call the proposed indicator the *t*-index. The *t*-index takes into account the number of citations, the number of coauthors on every published paper, and career duration. The proposed *t*-index makes the possible comparison of researchers at various stages of their careers.

The DEA model that we also proposed in the paper enabled us to directly compare considered scholars with their peers. In future research, it would be very interesting to test the DEA models based on various inputs and more than one output.

There are recent ideas to promote "download counts" of the published articles. This measure represents a new alternative for measuring the scientific output. Download counts enable measuring the article's impact immediately, in real-time. All proposed indices, including the proposed *t*-index, could be easily modified and include download counts instead of citation counts.